# Quantum Hall States Stabilized in Semi-magnetic Bilayers of Topological Insulators

R. Yoshimi[1+*], K. Yasuda[1+], A. Tsukazaki[2,3], K. S. Takahashi[4],

N. Nagaosa[1,4], M. Kawasaki[1,4] and Y. Tokura[1,4]

[1] *Department of Applied Physics and Quantum-Phase Electronics Center (QPEC),*

*University of Tokyo, Tokyo 113-8656, Japan*

[2] *Institute for Materials Research, Tohoku University, Sendai 980-8577, Japan*

[3] *PRESTO, Japan Science and Technology Agency (JST), Chiyoda-ku, Tokyo 102-0075,*

*Japan*

[4] *RIKEN Center for Emergent Matter Science (CEMS), Wako 351-0198, Japan.*

+These authors equally contributed to this work.

* Corresponding author: yoshimi@cmr.t.u-tokyo.ac.jp

**Abstract**

By breaking the time-reversal-symmetry in three-dimensional topological insulators with introduction of spontaneous magnetization or application of magnetic field, the surface states become gapped, leading to quantum anomalous Hall effect or quantum Hall effect, when the chemical potential locates inside the gap. Further breaking of inversion symmetry is possible by employing magnetic topological insulator heterostructures that host nondegenerate top and bottom surface states. Here, we demonstrate the tailored-material approach for the realization of robust quantum Hall states in the bilayer system, in which the cooperative or cancelling combination of the anomalous and ordinary Hall responses from the respective magnetic and non-magnetic layers is exemplified. The appearance of quantum Hall states at filling factor 0 and +1 can be understood by the relationship of energy band diagrams for the two independent surface states. The designable heterostructures of magnetic topological insulator may explore a new arena for intriguing topological transport and functionality.

**Introduction**

Three-dimensional (3D) topological insulator (TI) is a new class of material which possesses an insulating bulk with a two-dimensional (2D) Dirac electron state on its surface[1-4]. When the time-reversal-symmetry is broken with applying enough high magnetic fields or introducing spontaneous magnetization by doping magnetic impurities in 3D-TI, quantum Hall effect (QHE) or quantum anomalous Hall effect (QAHE) emerges as the hallmark of emergent states of 2D electron system[5]. Recently the QHE and QAHE have been observed in 3D-TI thin films[6,7] or cleaved bulk crystal[8] and their magnetically doped compounds[9-11]. In spite of the large energy gaps due to the Landau level (LL) formation of the Dirac state (~ 70 meV at $B = 14$ T)[12] or the presence of the spontaneous magnetization (~ 50 meV)[13], the observation of QHE or QAHE in the thin films has been limited so far at very low temperatures, typically below 100mK, probably because the magnetic impurities and crystalline imperfections make the quantization difficult due to the level broadening.

In this study, we propose a new magnetic TI system realizing the stable QH effect; the TI bilayer heterostructures composed of Cr-doped magnetic TI and pristine non-magnetic TI, hereafter referred to as semi-magnetic bilayers. From the theoretical[14] and experimental[10] points of view, the ground states of QHE and QAHE can be understood in the same context of topology. In the semi-magnetic TI bilayer, we can expect that the both magnetization *M* and magnetic field *B* identically drive the surface Dirac states in each TI layer to the QH states, which can be regarded as a hybrid phenomenon of QAHE in magnetic TI and QHE in non-magnetic TI layers with the common edge state. In other words, when the edge mode becomes stabilized, the system naturally converges to the QH state, yielding the quantized transverse conductivity $\sigma_{xy} \sim e^2/h$, irrespective of its origin. Moreover, as already suggested in the heterostructures with magnetic TI[15] or non-TI magnetism[16,17], the magnetic proximity effect on the non-magnetic TI surface from spatially separated magnetic layer may help the cyclotron gap or the exchange gap open much wider. Furthermore, the confinement of magnetic ions in a limited region of the heterostructure may suppress the disorder on the whole sample.

Results

**Transport properties of semi-magnetic bilayers** We fabricated TI bilayer heterostructures composed of $Cr_x(Bi_{1-y}Sb_y)_{2-x}Te_3$ (CBST) and $(Bi_{1-y}Sb_y)_2Te_3$ (BST) on semi-insulating InP(111) substrates using molecular-beam epitaxy (See Supplementary Table 1 for list of samples), as schematically illustrated in Figs. 1a and 1c. Here, we suppose the surface Dirac states should appear on the top surface of CBST as well as at the interface of BST and InP[12], as indicated by red arrows, but not for the interface of TIs[18]. First, we investigate the Bi/Sb composition ($y$) dependence of $R_{yx}$ in bare 5-nm CBST/5-nm BST heterostructures (Fig. 1a) with the same $y$ value to the both layers without AlO$_x$ gate dielectric. Figure 1b shows the Hall resistance $R_{yx}$ at $T = 2$ K under magnetic field $B$ for five samples with different $y$ values. While the anomalous Hall resistance at $B = 0$ T is positive for all the samples, the slope of $R_{yx}$ at high $B$ region (ordinary Hall component) changes its sign depending on $y$, suggesting that charge carrier type switches from hole to electron with decreasing $y$. These features resemble the case of the single layers of the BST and CBST, where it has been reported that Fermi level $E_F$ can be tuned by the $y$ value in BST and CBST compounds[19,20]. As for the observed $R_{yx}$ response, each BST and CBST layer appears to mainly contribute to the ordinary and anomalous Hall terms, respectively; a large value of 10 k$\Omega$ at $B = 14$ T is observed for $y = 0.88$, while a negative value for $y = 0.82$. From these investigations on the 5-nm/5-nm heterostructures, we concluded that the $y$ values of 0.86 and of 0.88 represent the system with the $E_F$ closest to the Dirac point and with the lowest hole density, respectively (see Supplementary Fig. 2).

Next, to finely tune the $E_F$ by electrical means, we defined a Hall-bar device for $y =$

0.88 with $AlO_x$ gate dielectric and Ti/Au gate electrode to fabricate field-effect transistor (FET) (see *Methods*). The schematic of vertical layered structure and top-view photographic image are shown in Fig. 1c and 1d, respectively. Note that the thickness of CBST layer ($t$ = 2 nm) was optimized so as to exhibit the largest $R_{yx}$ at high $B$ corresponding to the lowest hole density (see Supplementary Fig. 1). $R_{yx}$ and $R_{xx}$ in Fig. 1e under the transistor operation at $T$ = 0.5 K with $B$ = 0 T show systematic changes as a function of $V_G$; a single peak is observed at $V_G$ = 0.2 V where both $R_{yx}$ and $R_{xx}$ reach the maximum, manifesting that the Fermi energy is close to the Dirac point. In Fig.1f is shown the $B$ dependence of $R_{yx}$ measured with each $V_G$ application. The $R_{yx}$ slope at high $B$ region varies systematically from positive to negative with the change of $V_G$ from negative to positive, indicating the application of $V_G$ effectively tunes $E_F$ in the bilayer FET. As also shown in Fig. 1e and f, the anomalous Hall resistivity at $B$ = 0 is conspicuously enhanced at around $V_G$ = −1.3 V and 0.2 V; at these values of $V_G$ the clear quantized Hall plateau of $R_{yx}$ reaching $h/e^2$ = 25.8 kΩ is observed upon application of $B$ with the large contribution of anomalous Hall term. In addition, the coercive field in $R_{yx}$ hysteresis (see Supplementary Fig. 2) is always observed in the present $V_G$ range even in case of small $R_{yx}$ response, indicating that the ferromagnetism survives when $E_F$ is away from the Dirac point[9,10,20].

**Quantum Hall states stabilized in semi-magnetic TI bilayers** The feature of the QHE is also verified even at a higher temperature (e.g. 2 K) for both bare films without gate structure and FET device. The $B$ dependence of transverse and longitudinal conductivity $\sigma_{xy}$ and $\sigma_{xx}$ at $T = 2$ K for the bare bilayers (2-nm CBST/5-nm BST) of $y = 0.88$ (red) and 0.86 (blue) are displayed in Figs. 2a and 2b, respectively. For the $y = 0.88$ bilayer film, $\sigma_{xy}$ reaches the quantized value of $e^2/h$ accompanied by the decrease in $\sigma_{xx}$ towards 0 with increasing $B$, which are a clear indication of the QH state at the filling factor of $\nu = +1$. In the bilayer FET device of $y = 0.88$, the similar behavior of $\nu = +1$ QH state is observed at $V_G = -1.3$ V (red) as shown in Figs. 2c and 2d. In contrast, $\sigma_{xy}$ for the bare bilayer film of $y = 0.86$ and the $y = 0.88$ bilayer FET at $V_G = 1.17$ V show the asymptotic behavior towards 0 with increasing $B$, while $\sigma_{xx}$ decreases similarly to the case of the $\nu = 1$ QHE (blue). We attributed this to $\nu = 0$ QH state. In common for those two sample conditions, $E_F$ is located above the Dirac point where electrons are dominant conduction carriers. The contribution of negative ordinary Hall term apparently cancels out the anomalous Hall term, resulting in the $\nu = 0$ state emerging at high $B$. A notable feature of these bilayer films is that the QH states, both $\nu = 0$ and $\nu = +1$, are observable at 2 K, a much higher temperature than that of both QHE in BST[7,8] and QAHE in CBST single layer films[9-11]; this is true also in the case of bare films without $E_F$ fine tuning.

**Discussion**

The emergence of two QH states at $\nu = 0$ and +1 can be understood by the relationship of energy diagrams between the two independent surface bands of BST and CBST shown in Figs. 2e and 2f. We first focus on the energy relationship between the magnetization-induced gap on the top surface of CBST layer and the Dirac point on the bottom surface of BST layer. When $V_G$ of 0.2 V is applied, $E_F$ locates around the center of the gap of CBST surface state, since anomalous Hall term in $R_{yx}$ at $B = 0$ T reaches maximum, as shown in Fig. 1e. With applying magnetic field, the ordinary Hall term is added to the anomalous Hall term. The observed positive ordinary Hall term at $V_G = 0.2$ V shown in the center panel of Fig. 1f indicates that $p$-type carrier dominantly comes from the bottom BST surface state because the top CBST surface state with $E_F$ within the gap should minimally contribute to the ordinary Hall term. Therefore, the relative energy position around the Dirac point between CBST and BST surfaces at $V_G = 0.2$ V under $B = 0$ T is as depicted in Fig. 2e. Under high magnetic field, the Landau levels (LLs) are formed from the Dirac band dispersion as schematically illustrated in Fig. 2f. Here, $n = 0$ LL forms on one side of the massive Dirac cone depending on a sign of the mass term of TI. In case of CBST, it is known to form at the bottom of higher energy one[20]. When $E_F$ is below the $n = 0$ LLs of both top and bottom surfaces, the $\nu = +1$ QH state emerges. Following the band relationship shown in Fig. 2f, fine tuning of $E_F$ between the two $n = 0$ LLs enables us to achieve the $\nu = 0$ state.

To identify $E_F$ location in the two surface bands drawn in Figs. 2e and 2f more precisely, $V_G$-control measurements for $R_{xx}$ and $R_{yx}$ are performed. The QHE in FET is clearly demonstrated at various magnetic fields in Figs. 3a-d. Hallmarks of $\nu = +1$ are observed at around $V_G = -1.3$V (red arrow in Fig. 3b) with increasing $B$; $R_{yx} \sim 25.8$ k$\Omega$ (Fig. 3a), $\sigma_{xy} = + e^2/h$ (Fig. 3c), and $R_{xx}$ and $\sigma_{xx} \sim 0$ (Figs. 3b and 3d). The other QH plateau at $\nu = 0$ in $\sigma_{xy}$ (Fig. 3c) is realized around $V_G = 1.1$ V; $R_{yx} \sim 0$ (Fig. 3a) and high $R_{xx}$ (blue arrow in Fig. 3b). Correspondingly, $R_{xx}$ in the inset of Fig. 3b increases up to about 100 k$\Omega$ with increasing $B$ to 14 T. In this case, positive anomalous and negative ordinary Hall resistivity almost cancels out, resulting in a small value of total $R_{yx}$. In addition, a dip in $\sigma_{xx}$ vs $V_G$ curve is simultaneously obtained at $V_G = 1.1$ V in Fig. 3d. This exemplifies the $\nu = 0$ pseudo-spin Hall insulator state[7], in which we view the top and bottom degrees of freedom in the surface states as the pseudo spins, as expected from band diagram shown in Fig. 2f.

The appearance of $\nu = 0$ in $\sigma_{xy}$ at high $B$ between two peaks of $\sigma_{xy}$ at $B = 0$ T (black curve in Fig. 3c) is a compelling evidence for the $E_F$ location as discussed above. Indeed, the QH states $\nu = 0$ and $+1$ can be clearly resolved in the conductivity mapping ($\sigma_{xy}$ ($V_G$), $\sigma_{xx}$ ($V_G$)) at various $V_G$, as shown in the inset of the Fig. 3d.

Finally, we discuss why the QH state in the bilayers of BST and CBST are observable up to higher temperatures or much more stable than those of single-layer films of BST and CBST. We raise two possible reasons; the one is a magnetic proximity effect on the BST surface from the adjacent ferromagnetic CBST layer and the other is less disorder at the surface conduction channels in the present bilayer setup. Regarding the proximity effect, the increase in Hall response with application of magnetic field at negative $V_G$ (hole accumulation) is approximately three times as large as that at positive $V_G$ (electron side) (see Supplementary Fig. 3). This enhancement of Hall response at negative $V_G$ coincides with the peak of anomalous $\sigma_{xy}$ response ($B = 0$ T, black curve in Fig. 3c) at $V_G = -1.3$ V, pointing to the proximity effect from the magnetic CBST layer to the BST surface state (see also Supplementary Note 3). Second, as for the disorder, the interface combinations of AlO$_x$/CBST and BST/InP may have less disorder than those of AlO$_x$/BST and CBST/InP; this is speculated from the small Hall response of the semi-magnetic heterostructure with the inverted structure of BST/CBST/InP (see Supplementary Fig. 4). These two effects may work cooperatively to stabilize the QHE states in the semimagnetic bilayers.

In conclusion, we have successfully resolved the QH states in semi-magnetic bilayers of topological insulator. The surface state readily exhibits the $\nu = 0$ and $+1$ QH states under a relatively small magnetic field with a large contribution from the anomalous Hall term in magnetic TI layer. These QH states are accounted for in terms of the magnetization-induced gap and/or the formation of LLs at each component-layer surface state. Furthermore, the observation at a relatively high temperature ($T = 2$ K) suggests the semi-magnetic structure may have the proximity effect while suppressing the disorder effect for surface transport. TI-based semi-magnetic heterostructures and superlattices may provide a new platform in exploring new functionality and exotic phases of TIs[21,22].

**Methods**

**MBE thin film growth**    Bilayer thin films of $(Bi_{1-y}Sb_y)_2Te_3$ (BST) and $Cr_x(Bi_{1-y}Sb_y)_{2-x}Te_3$ (CBST) were fabricated by molecular-beam epitaxy (MBE) on semi-insulating InP (111) substrates. The Bi/Sb composition ratio $y$ for bare films was calibrated by the beam equivalent pressure of Bi and Sb, for example $6 \times 10^{-7}$ Pa and $4.4 \times 10^{-6}$ Pa for $y = 0.88$. Since the formation of $AlO_x$/Ti/Au gate structure changes the $E_F$ position in the channel, we fabricated FETs with various $y$ values close to 0.88 for maximizing $R_{yx}$ at $V_G = 0$ V under large $B$. The designed $y$ value for the FET channel was 0.84, but from the comparison with the transport properties of the bare films the $y$ value is indicated as 0.88 in this paper for the sake of consistency and readability. The Cr concentration $x$ was determined by the flux ratio of Cr / (Bi + Sb). The Te flux was overs-supplied with keeping the Te / (Bi + Sb) ratio at about 20. Prior to the growth of the first layer of BST, we started with supplying Te and Sb for a growth of monolayer $Sb_2Te_3$ buffer layer to construct a smooth interface with InP substrate.

**FET device fabrication**    After the epitaxial growth of BST layer, films were annealed *in-situ* at 380°C to make the surface smoother under the exposure of Te flux. The same procedure was employed for the following CBST layer. For the preparation of FET devices,

AlO$_x$ capping layer was deposited at room temperature with an atomic layer deposition system immediately after the discharge of the samples from MBE. The device pattern was defined by photolithography and Ar ion-milling processes.

**Transport measurements**   Ohmic-contact electrodes and top gate electrode were Ti/Au deposited with an electron-beam evaporator. Transport measurements for bare films were conducted using the d.c. transport option of physical property measurement system (PPMS) by Quantum Design. FET devices were measured in PPMS with employing a lock-in technique at a frequency (~ 3 Hz) and with a low excitation current (~ 10 nA) to suppress heating effect. A series resistance of 100 MΩ was introduced to maintain a constant current condition. Low temperature (< 2 K) measurements were performed using the $^3$He option of PPMS.

**Acknowledgements**

We thank Y. Kozuka, J. Falson, J. G. Checkelsy and Y. Takaoka for fruitful discussions and experimental contributions. R. Y. is supported by the Japan Society for the Promotion of Science (JSPS) through a research fellowship for young scientists. This research was supported by the Japan Society for the Promotion of Science through the Funding Program for World-Leading Innovative R & D on Science and Technology (FIRST Program) on "Quantum Science on Strong Correlation" initiated by the Council for Science and Technology Policy and by JSPS Grant-in-Aid for Scientific Research(S) No.24224009 and 24226002.

**Author contributions**

R. Y. and K. Y. performed thin film growth and device fabrication. The transport measurements and data analysis for bare bilayer films and FET devices were conducted by R. Y. and K. Y., respectively. R. Y., K. Y., A. T. and Y. T. wrote the manuscript with contributions from all authors. A. T., K. S. T., N. N., M. K. and Y. T. jointly discussed the results and guided the project. Y. T. conceived and coordinated the project.

**Competing Financial Interests**

The authors declare no competing financial interests.

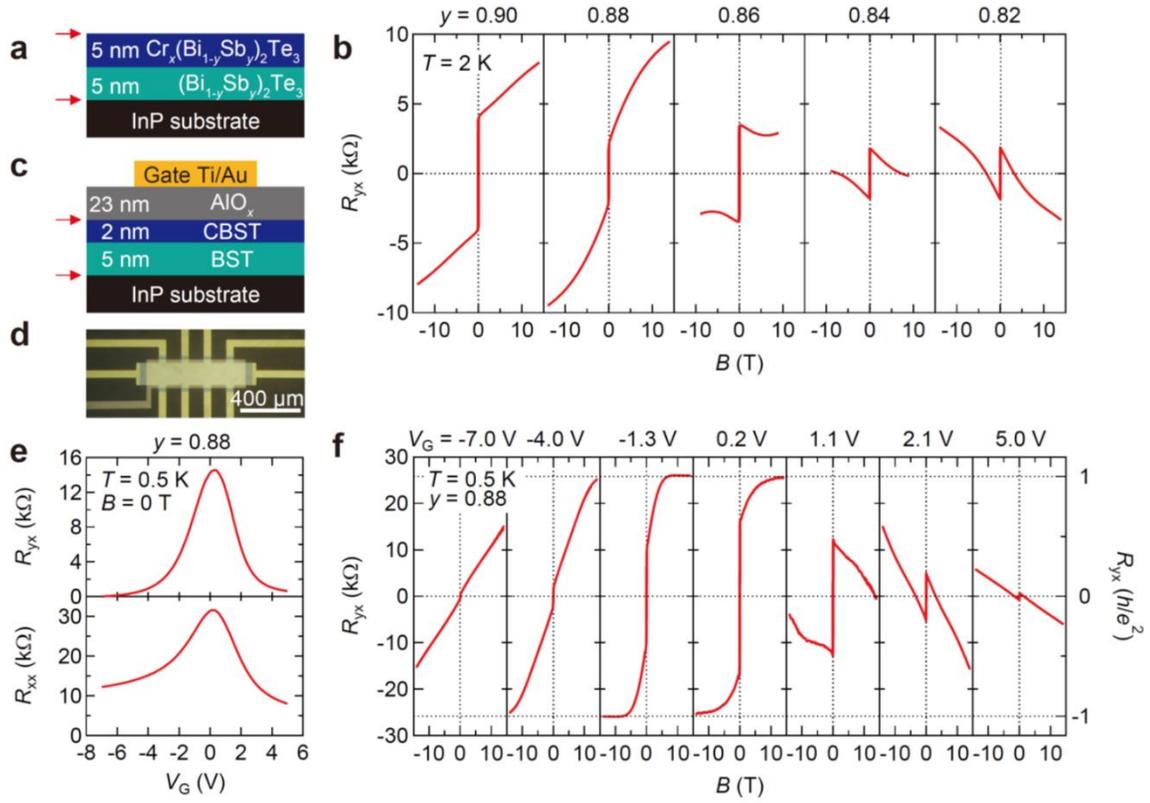

**Figure 1 | Hall responses $R_{yx}$ in $Cr_x(Bi_{1-y}Sb_y)_{2-x}Te_3/(Bi_{1-y}Sb_y)_2Te_3$ semi-magnetic TI bilayers. a,** A schematic of semi-magnetic TI bilayer composed of 5-nm CBST/5-nm BST. CBST and BST represent the $Cr_x(Bi_{1-y}Sb_y)_{2-x}Te_3$ and $(Bi_{1-y}Sb_y)_2Te_3$, respectively. Arrows indicate the interfaces where the Dirac state exists. **b,** Transverse resistivity $R_{yx}$ as a function of magnetic field $B$ at $T = 2$ K for several bare bilayers with different $y$. **c, d,** Cross-sectional schematic (c) and top-view photograph (d) of a field effect transistor (FET) with a Hall-bar channel of 2-nm CBST/5-nm BST ($y = 0.88$). **e,** $V_G$ dependence of $R_{yx}$ and longitudinal resistivity ($R_{xx}$) at $B = 0$ T. **f,** Magnetic field dependence of $R_{yx}$ at $T = 0.5$ K for

several gate voltage $V_G$ for FET.

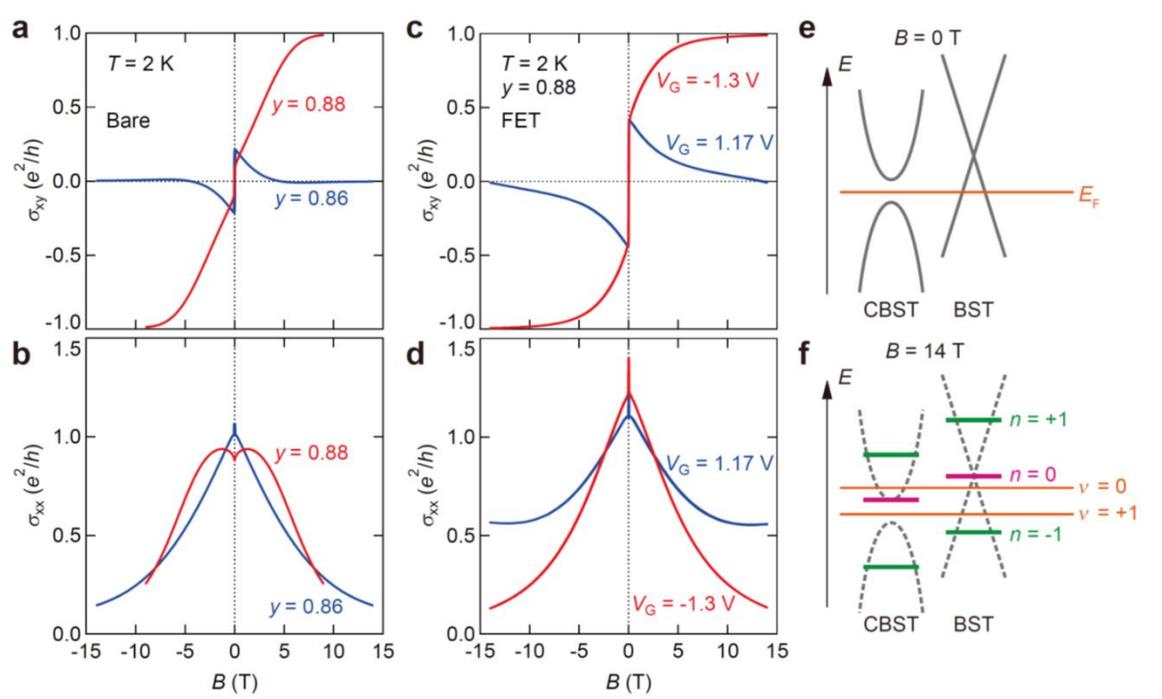

**Figure 2 | Conductivity responses in semi-magnetic TI bilayers observed at $T = 2$ K. a, b,** Magnetic field dependence of longitudinal and transverse conductivity $\sigma_{xy}$ (a) and $\sigma_{xx}$ (b) for $y = 0.88$ and $y = 0.86$ bare films of 2-nm CBST/5-nm BST at $T = 2$ K. **c, d,** Magnetic field dependence of $\sigma_{xy}$ (c) and $\sigma_{xx}$ (d) for the FET device of $y = 0.88$ at $V_G = -1.3$ V ($\nu = +1$) and $V_G = 1.17$ V ($\nu = 0$). **e, f,** Schematic band diagram for the surface states of top CBST and bottom BST layers at magnetic field $B = 0$ T and 14 T, respectively. $E_F$ represents the Fermi level at $V_G = 0.2$ V for the $y = 0.88$ FET. In (f), Landau levels (LLs) $n = +1, 0$ and $-1$ are denoted by horizontal lines. Filling factor $\nu$ is indicated when $E_F$ locates at the depicted energy position.

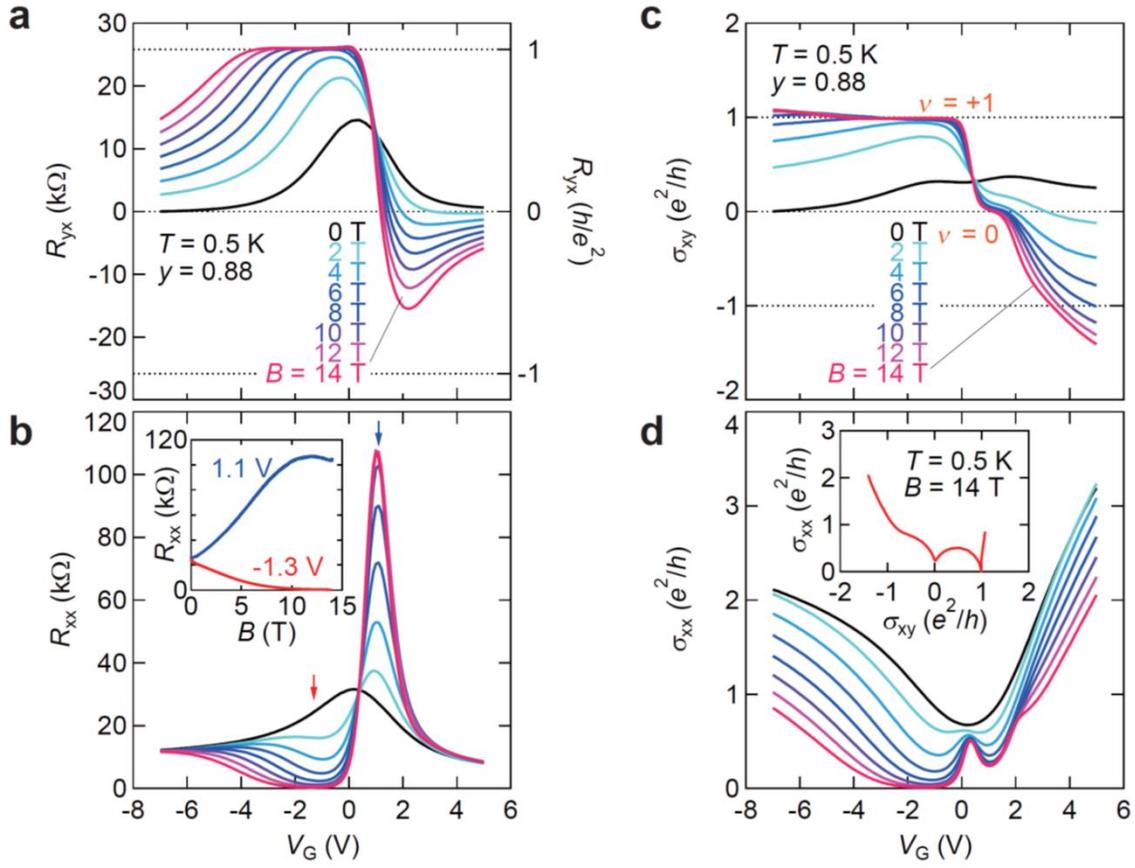

**Figure 3 | Magnetic field dependence of QH states in gate-tuned semi-magnetic TI bilayers. a, b,** $V_G$ dependence of $R_{yx}$ (a) and $R_{xx}$ (b) for $y = 0.88$ FET at $T = 0.5$ K under various magnetic fields. The inset in b shows the magnetic field dependence of $R_{xx}$ at $V_G = 1.1$ V (blue) and $V_G = -1.3$ V (red). Blue and red arrows in the main panel of b represent gate voltages of $V_G = 1.1$ V and $-1.3$ V, respectively. **c, d,** $V_G$ dependence of $\sigma_{xy}$ (c) and $\sigma_{xx}$ (d) at $T = 0.5$ K under various magnetic fields. The inset in d plots the ($\sigma_{xy}(V_G)$, $\sigma_{xx}(V_G)$) at various $V_G$ under magnetic field $B = 14$ T.